\documentstyle[12pt]{article}

\newcommand{\mysection}{\setcounter{equation}{0}\section}

\begin{document}
\hfill{FERMILAB-Pub-93/270-T} 
\vskip 0.1cm
\hfill{ITP-SB-93-55} 
\vskip 0.1cm
\hfill{THU-93/23} 
\vskip 0.1cm
\vskip 0.5cm
\centerline{\large\bf { Top Quark Production Cross Section}}
\vskip 0.5cm
\centerline{\sc E. Laenen,}
\vskip 0.3cm
\centerline{\it Fermi National Accelerator Laboratory,} 
\centerline{\it P.O. Box 500, MS 106}
\centerline{\it Batavia, Illinois 60510}
\vskip 0.3cm
\centerline {\sc J. Smith,\footnote{On leave of absence from the
Institute for Theoretical Physics, SUNY at Stony Brook, NY 11794-3840}}
\vskip 0.3cm
\centerline{\it Institute for Theoretical Physics,}
\centerline{\it University of Utrecht,}
\centerline{\it P.O.B. 80006, 3508 TA, Utrecht,}
\centerline{\it The Netherlands,}
\vskip 0.3cm 
\centerline{and}
\vskip 0.3cm 
\centerline{\sc W. L. van Neerven}
\vskip 0.3cm 
\centerline{\it Instituut Lorentz,}
\centerline{\it University of Leiden,}
\centerline{\it P.O.B. 9506, 2300 RA, Leiden,} 
\centerline{\it The Netherlands.}
\vskip 0.3cm
\centerline{August 1993}
\vskip 1.2cm
\centerline{\bf Abstract}
\vskip 0.3cm

The production rate  for top quarks at the Fermilab
Tevatron is presented using the exact 
order $\alpha_s^3$ corrected cross section and the
resummation of the leading soft gluon corrections
in all orders of perturbation theory.
\vfill
\newpage

%------------------This is Section 1---------------------------------
\mysection{Introduction}
%----------------------------------------------------------

At the Tevatron, the top quark will be mainly
produced through $t\bar{t}$ pair creation. Both
top quarks will then decay to $(W,b)$ pairs, and then
each $W$ can decay either hadronically or leptonically.
It is in the channel where both $W$'s decay leptonically,
one to a $(e,\nu_e)$ pair, the other to a 
$(\mu,\nu_{\mu})$ pair, that a large part of the current 
search effort to find the top quark 
is concentrated. This is because
the background in this channel from $W^+W^-$ plus
jets is fairly small. Using the known branching fractions
of the above decays, and taking acceptances into account,
an experimental top production cross section
is determined, which is then compared with a curve of the theoretical
cross section as a function of the top quark
mass $m$. At present
the top quark has not yet been discovered at the
Tevatron, and thus an estimate of the theoretical
cross section is needed to either determine its mass
or establish a lower limit on it. 
We therefore discuss the estimates of this cross section here.

There have been previous predictions of the top quark cross
section based on the results of the fixed order 
$\alpha_s^2$ plus $\alpha_s^3$
contributions in perturbative QCD
(pQCD) \cite{NDE},\cite{Altetal}, \cite{Betal},\cite{Ellis}.
As with all fixed order pQCD calculations, these contain
a scale (factorization scale = renormalization scale) which
reflects the size of the uncalculated $O(\alpha_s^4)$ and higher order
terms. Although the dependence on this scale is relatively
flat, indicating that the present result is stable under scale changes, 
the size of the $O(\alpha_s^3)$ term is disturbing.
In fig.1 we show the ratio
$R = (\sigma_{\rm exact}^{(0)} + \sigma_{\rm exact}^{(1)})/
\sigma_{\rm exact}^{(0)}$
for top quark masses in the range relevant to the Fermilab
Tevatron. With $\sigma_{\rm exact}^{(n)}$ we denote the 
exact order $\alpha_s^{(n+2)}$ contribution to the cross section,
which implies that $\sigma_{\rm exact}^{(0)}$ stands for the Born 
contribution.
Here, and throughout this paper, we have
used recent parton distribution functions (MRSD\_') \footnote{We
thank W.J. Stirling for sending us the DIS scheme MRS parton
distribution funtions.}
\cite{MRS}, choosing the DIS factorization scheme, and the two-loop
running coupling constant (in the $\overline{\rm MS}$ 
scheme) with five active flavours
and $\Lambda = 0.152$ GeV. The ratio $R$ of the next-to-leading order
(NLO) to the leading-order (LO) term is usually referred to as the 
K-factor.

One notes from fig.1 that the higher order corrections
in the $q\bar{q}$ channel are small, whereas those in 
the $gg$ channel are 70\% or larger. 
Throughout the mass range we consider 
the $qg$ and $\bar{q}g$ channel give negligible 
contributions so we
do not consider them.  To show which channel
has the largest cross section in this mass range we plot
the ratios of the NLO $gg$ and $q\bar{q}$ contributions to the 
total result in fig.2. Here we see that as the top quark 
mass increases the $q\bar{q}$ channel contribution is larger than the
$gg$ channel one, so the effect of the large K-factor
in the latter channel, as seen in fig.1,
decreases. However, at masses around 150 ${\rm GeV}/c^2$
the $gg$ channel still contributes around 20\%. This is enough
to worry about even higher order corrections. 

These large corrections are predominantly from the threshold
region for heavy quark production, where we have shown 
previously that initial state gluon bremsstrahlung (ISGB)
is responsible for the large corrections at NLO \cite{MSSN}.
To demonstrate this we have run our cross sections programs
with a cut on the variable
$\eta = (\hat{s}-4m^2)/4m^2$, where $\hat{s}$
is the square of the parton-parton cms energy so that:
\begin{equation}
\sigma(\eta_{cut}) = \sum_{i,j}\int_0^{\eta_{cut}} d\eta 
\frac{1}{1+\eta}\; \Phi_{ij}(\eta,\mu^2)
\;\hat{\sigma}_{ij}(\eta,m^2,\mu^2), \label{one}
\end{equation}
where $\Phi_{ij}$ denotes the parton flux and
$\hat{\sigma}_{ij}$ the partonic cross section
for the incoming partons $i$ and $j$. The precise definitions of 
these functions are given in Ref.\cite{MSSN}. The factorization
scale is $\mu$, which we have set equal to the renormalization
scale. Notice that the maximum
value of $\eta_{cut}$ is given by $\eta_{max}=(S-4m^2)/4m^2$, where
$S$ denotes the square of the total hadronic energy in the cms system.
As we increase
$\eta_{cut}$ from small values ($\simeq 10^{-1}$) to larger values,
where the actual value of the cross section is approached,
there is a rapid rise in $\sigma(\eta_{cut})$. Figures
3 and 4 show $\sigma(\eta_{cut})$ for the $gg$ and the $q\bar{q}$
contributions to the cross section. 
The fact that both cross sections
rise sharply around $\eta_{cut}=1$ (where $\hat{s}\approx 8m^2$)
indicates that the threshold region is very important. This 
is especially true in the $gg$ channel, which dominates
the total cross section at smaller top quark masses.
Both figures also show that the cross 
section flattens out if $\eta_{cut}$ is increased further,
indicating that partonic processes with $\hat{s} >> 4m^2$
contribute very little to the cross section.

In a previous paper \cite{LSN} we carefully examined the dominant logarithms
from ISGB which are the cause of the large corrections
near threshold. Such logarithms have been studied previously
in Drell-Yan (DY) \cite{DY} production at fixed target energies 
(again near threshold) where they are responsible for
correspondingly large corrections.  In \cite{LSN} we exploited the 
analogy between DY and heavy quark production cross sections and proposed
a formula to resum the leading and next-to-leading logarithms
in pQCD to all orders. Since the contributions due to these
logarithms are positive (when $\mu=m$),
the effect of summing the higher order corrections
increases the top quark production cross section over that
predicted in $O(\alpha_s^3)$. 
This sum, which will be indentified with $\sigma_{\rm res}$, depends
on a nonperturbative parameter $\mu_0$. The reason that
a new parameter has to be introduced is due to the fact that
the resummation is sensitive to the scale at which
pQCD breaks down. As we approach the threshold region
other, nonperturbative, physics plays a r\^ole (higher twist,
bound states, etc) indicated by a dramatic increase in 
$\alpha_s$. We chose to simply stop the resummation
at a specific scale $\mu_0$ where 
$\Lambda << \mu_0 << m$ since it is not
obvious how to incorporate the nonperturbative physics.
Note that our resummed corrections diverge for small $\mu_0$ 
but this is {\em not} physical since they should be
joined smoothly onto some nonperturbative prescription
and the total cross section will be finite. However, at the moment
our total resummed corrections depend on the parameter
$\mu_0$ for which we can only make a rough estimate.
See \cite{LSN} for more details.

Let us begin by showing the effects of including only
the leading soft gluon contribution at $O(\alpha_s^4)$,
which we call $\sigma^{(2)}_{\rm app}$ .
An explicit expression for this contribution 
is given in \cite{LSN}.
Here $\sigma^{(2)}_{\rm app}$ stands for the 
approximation to $\sigma^{(2)}_{\rm exact}$ where only
the leading soft gluon corrections are taken into
account. 
Figure 5 shows three curves for the exact cross section
calculated through $O(\alpha_s^3)$
at the scales $\mu = 2m, m$ and $m/2$. 
This is the traditional method of estimating the size
of uncalculated higher order contributions. 
For comparison
we add to the $\mu = m$ case the approximate $O(\alpha_s^4)$ 
contribution, yielding a total cross section which is 
not in the range spanned by the previous curves, but 
slightly above. 
Therefore, the traditional method does not work very
well in this case, due to the size of the corrections, 
as was already pointed out in \cite{BTG}.

Now we study the effect of the resummation, which 
depends on $\mu_0$, by calculating $\sigma_{\rm res}$.  
However, because we know the exact $O(\alpha_s^3)$ result, 
we can make an even better estimate of the cross
section by calculating the quantity 
\begin{equation}
\sigma_{\rm imp}
= \sigma_{\rm res} - \sigma^{(1)}_{\rm app}
+ \sigma^{(1)}_{\rm exact}, \label{two}
\end{equation}
which we call the improved cross section.  We remind the 
reader that $\sigma^{(n)}$ denotes the $O(\alpha_s^{(n+2)})$ 
contribution to the cross section. 
Further $\sigma_{\rm exact}^{(n)}$ denotes
the exact calculated cross section and 
$\sigma_{\rm app}^{(n)}$ the approximated
one where only the leading soft gluon corrections
are taken into account. We compare the
results for $\sigma_{\rm imp}$ versus the 
fixed order result $\sigma_{\rm exact}^{(0)} + \sigma_{\rm exact}^{(1)}
+ \sigma^{(2)}_{\rm app}$ in fig.6 for various
(reasonable) values of $\mu_0$. Note that $\mu_0$
need not be the same in the $q\bar{q}$ 
and $gg$ channels because the convergence properties
of perturbation series could be different 
in these channels and depend
on the factorization scheme. For example the
cross section due to the $q\bar{q}$ process seems to converge faster. 

As we decrease $\mu_0$ the cross sections increase.
In figure.6 we show three curves for various choices
of $\mu_0$. 
The requirement that $\Lambda << \mu_0 << m_t$ 
is satisfied by all three choices. This is not
a very restrictive requirement in the
sense that it still leaves a large range
of values possible, thus rendering the 
cross section from pQCD more uncertain than
thought previously. 

In fig.7 we show the contribution from the 
$q\bar{q}$ channel to fig.6, while the
$gg$ channel is given in fig.8. The plots
show that the range of possible cross sections
is quite narrow for the $q\bar{q}$ channel
(due to the smaller relative correction) while
it is relatively large in the $gg$ channel.
The curves are calculated in the DIS factorization
scheme, where the corrections are smaller because
most of them have been absorbed in the
parton distribution functions. 
Thus the resummation is successful for the
$q\bar{q}$ channel, in the sense that
$\sigma_{\rm imp}$ differs very little
from $\sigma_{\rm exact}^{(0)}+\sigma_{\rm exact}^{(1)}$.
This is unfortunately not the case for $gg$.
We have also checked that these results change very little
when using CTEQ \cite{CTEQ} parton distribution functions.
This is to be expected because, due to the fact that top is so heavy, 
the cross section is mainly sensitive
to parton distribution functions at large $x$
where they have been well measured.

Finally in Table 1 we present a lower limit
estimate (from \cite{Lae}), and a central value
and upper limit estimate. The latter two are also shown 
in fig.6 as the central and upper solid line respectively,
and are obtained as follows. For both estimates 
we used the improved cross section (\ref{two}). Each
of the three terms in (\ref{two}) was calculated
according to (\ref{one}), with $\eta_{cut} = \eta_{max}$. 
Furthermore, as stated earlier, we used the MRSD\_'
parton distribution functions, and chose the DIS 
factorization scheme and the two-loop
running coupling constant (in the $\overline{\rm MS}$
scheme) with five active flavours
and $\Lambda = 0.152$ GeV. The exact
partonic cross sections used to calculate
$\sigma_{\rm exact}^{(0)}+\sigma_{\rm exact}^{(1)}$
were obtained from
\cite{NDE, Altetal, Betal}. The approximate 
DIS scheme partonic cross section 
used for determining $\sigma_{\rm app}^{(1)}$
is given explicitly in eqn. (2.10) in \cite{LSN}.
For the partonic resummed cross section we used
eqn. (3.24) in the same reference. For all three cross section 
contributions we chose $\mu=m$. However, the 
central value and upper limit estimates differ
by the value chosen for the nonperturbative parameter
$\mu_0$. Thus, for the central value we 
chose $\mu_0 = 0.1m$ and $\mu_0 = 0.25m$ for the $q\bar{q}$
and $gg$ channels channels respectively, whereas for the upper
limit we chose $\mu_0 = 0.05m$ and $\mu_0 = 0.2m$,
respectively.
It should be noted that these
estimates merely represent what we think are  
reasonable choices for the parameter $\mu_0$,
and are thus still not completely rigorous.
As far as the lower limit estimate is concerned,
note that we can write
\begin{equation}
\sigma_{\rm imp} = \sigma_{\rm exact}^{(0)} + \sigma^{(1)}_{\rm exact}
+ \sum_{i=2}^{\infty} \sigma^{(i)}_{\rm app}, \label{three}
\end{equation}
Since $\sigma^{(i)}_{\rm app} > 0$ for all $i$ at $\mu = m$ \cite{LSN},
the true total cross section is likely larger than
\begin{equation}
\sigma_{\rm lower} = \sigma_{\rm exact}^{(0)} + 
\sigma^{(1)}_{\rm exact} + \sigma^{(2)}_{\rm app},
\label{four}
\end{equation}
so we are justified in using this value as a lower limit.
The calculation of the three terms in (\ref{four}) goes
analogously as described above for the central value and
upper limit cases. It differs only in the fact that
we do not have the nonperturbative parameter $\mu_0$
here
and that we used the conservative value $\Lambda = 0.105$ GeV in 
the expression for $\alpha_s$ when we make this estimate for  
the lower limit. The explicit expression for 
$\sigma^{(2)}_{\rm app}$ in the DIS scheme is given in eqn. (2.14)
in \cite{LSN}. 

In conclusion we have demonstrated that the leading and next-to-leading
logarithmic corrections to the top quark production cross section
are large near threshold and have to be resummed to give a more precise estimate
for the cross section. When the ISGB contributions are resummed to all orders 
in pQCD a new scale $\mu_0$ has to be introduced, which measures the sensitivity of
the cross section to nonperturbative physics. 
The top quark production cross section at the Fermilab Tevatron is
sensitive to this new scale, mainly via the contribution from the 
gluon-gluon fusion channel.

{\bf Acknowledgements}

The work in this paper was supported in part under the
contracts NSF 93-09888 and DOE DE-AC02-76CH03000.
Financial support was also provided by the Texas National
Research Laboratory Commission.

\newpage
\vspace{-1cm}

\hspace{-0.1cm}\begin{tabular}{||c||c|c|c||c||c|c|c||} \hline 
{\rm m$_{\rm top}$} & $\sigma$ (pb) & $\sigma$ (pb) & $\sigma$ (pb) &
{\rm m$_{\rm top}$} & $\sigma$ (pb) & $\sigma$ (pb) & $\sigma$ (pb) \\
{}     & Lower     & Central  & Upper &
{}     & Lower     & Central  & Upper  \\ \hline
90       & 148   &180  &259  &  146     & 12.1 &13.6  &16.2  \\
92       & 132   &160  &227  &  148     & 11.3 &12.6  &15.0  \\
94       & 118   &143  &204  &  150     & 10.5 &11.7  &13.8  \\
96       & 106   &127  &180  &  152     & 9.79 &10.9  &12.8  \\
98       & 95.2  &114  &158  &  154     & 9.14 &10.1  &11.9  \\
100      & 86.3  &102  &141  &  156     & 8.52 &9.40  &11.0  \\
102      & 77.8  &92.4  &127  &  158     & 7.94 &8.77  &10.3  \\
104      & 70.6  &83.2  &113  &  160     & 7.41 &8.16  &9.53  \\
106      & 64.0  &75.4  &102  &  162     & 6.92 &7.62  &8.82  \\
108      & 58.1  &68.0  &90.9  &  164     & 6.48 &7.11  &8.25  \\
110      & 52.7  &61.6  &81.4  &  166     & 6.07 &6.67  &7.70  \\
112      & 48.2  &55.9  &73.6  &  168     & 5.68 &6.23  &7.18  \\
114      & 43.9  &51.2  &66.6  &  170     & 5.32 &5.83  &6.68  \\
116      & 40.2  &46.6  &60.6  &  172     & 4.98 &5.45  &6.25  \\
118      & 36.8  &42.4  &54.7  &  174     & 4.67 &5.10  &5.83  \\
120      & 33.7  &38.9  &49.7  &  176     & 4.38 &4.79  &5.46  \\
122      & 31.1  &35.6  &45.4  &  178     & 4.11 &4.49  &5.09  \\
124      & 28.4  &32.6  &41.1  &  180     & 3.86 &4.21  &4.78  \\
126      & 26.2  &29.9  &37.5  &  182     & 3.63 &3.94  &4.47  \\
128      & 24.2  &27.5  &34.5  &  184     & 3.40 &3.70  &4.16  \\
130      & 22.3  &25.4  &31.6  &  186     & 3.20 &3.48  &3.92  \\
132      & 20.6  &23.3  &29.0  &  188     & 3.00 &3.27  &3.67  \\
134      & 19.1  &21.5  &26.5  &  190     & 2.83 &3.06  &3.44  \\
136      & 17.6  &19.9  &24.3  &  192     & 2.67 &2.88  &3.22  \\
138      & 16.3  &18.3  &22.4  &  194     & 2.50 &2.70  &3.02  \\
140      & 15.1  &16.9  &20.5  &  196     & 2.36 &2.55  &2.85  \\
142      & 14.0  &15.7  &19.0  &  198     & 2.22 &2.40  &2.68  \\
144      & 13.0  &14.5  &17.4  &  200     & 2.09 &2.26  &2.52  \\ \hline
\end{tabular}
\vspace{0.4cm}

Table 1. The first and fifth column contain the top quark mass 
in ${\rm GeV}/c^2$. The columns denoted by `Lower' show 
our lower limit estimate
of the top quark cross section in picobarns, the columns denoted
by `Central' show our central value estimate, and the columns denoted
by `Upper' show our upper limit estimate.

%----------------------------References-------------------------------------
%

\newpage

\centerline{\bf \large{Figure Captions}}

\begin{description}
\item[Fig. 1.]
The ratio $R$ (`K-factor')
for the NLO exact top quark cross section as a function
of the top quark mass. Plotted are the K-factors for the
total cross section (solid line), and individually for 
the $q\bar{q}$ channel 
(long-dashed line) and $gg$ channel (short-dashed line).
\item[Fig. 2.]
Fraction of $q\bar{q}$ channel (long-dashed line)
and $gg$ channel (short-dashed line) contribution to total NLO
cross section as function of the top quark mass.
\item[Fig.3]
Cross section as function of $\eta_{cut}$ (see eq.(1))
for $q\bar{q}$ channel. We used the DIS scheme
MRSD\_' parton distribution functions, and
$m = 100$ ${\rm GeV}/c^2$. Plotted are 
$\sigma(\alpha_s^2)$ (solid line) and
$\sigma(\alpha_s^3)$ (dashed line).
\item[Fig.4]
Same as fig.3 but now for the $gg$ channel.
\item[Fig.5]
The NLO exact cross section as a function
of the top quark mass for three choices of 
scale: $\mu = m/2$ (upper solid line), $\mu = m$
(central solid line) and $\mu = 2m$ (lower solid line), 
and the NLO exact cross section 
plus the $O(\alpha_s^4)$ contribution at $\mu = m$,
(dashed line). 
\item[Fig.6]
The $O(\alpha_s^4)$ cross section at $\mu = m$ (dashed line)
and $\sigma_{\rm imp}$ (eq. (2)) for three choices
of scale $\mu_0$, the two numbers
per line corresponding to the $q\bar{q}$ and $gg$ channels
respectively: 0.05 $m$/0.2 $m$ (upper solid line), 
0.1 $m$/0.25 $m$ (central solid line), 
0.2 $m$/0.3 $m$ (lower solid line).
\item[Fig.7]
Range of cross sections for $q\bar{q}$ channel only.
The two solid lines span the exact NLO cross section
for the range $m/2 < \mu < 2m$, and the two
dashed lines $\sigma_{\rm imp}$ for the range
$0.05 m < \mu_0 < 0.2 m$.
\item[Fig.8]
Range of cross sections for $gg$ channel only. 
The two solid lines span the exact NLO cross section
for the range $m/2 < \mu < 2m$, and the two
dashed lines $\sigma_{\rm imp}$ for the range
$0.2 m < \mu_0 < 0.3m$.
\end{description}

\end{document}